\documentclass[11pt, a4paper]{article}
\usepackage{jheppub}
\usepackage{graphicx}
\usepackage{epsfig,amsmath,amsthm}
\usepackage{epstopdf}

\newcommand{\be}{\begin{equation}}
\newcommand{\ee}{\end{equation}}
\newcommand{\bea}{\begin{eqnarray}}
\newcommand{\eea}{\end{eqnarray}}
\def\bse{\begin{subequations}}
\def\ese{\end{subequations}}

\def\IZ{\relax\ifmmode\hbox{Z\kern-.4em Z}\else{Z\kern-.4em Z}\fi}

\newcommand{\non}{\nonumber \\}

\def\half{\frac{1}{2}} 

\def\del{{\partial}}

\def\hd{{\hat d}}

  \def\eps{\epsilon}

\def\presub{\vspace{.5cm} \noindent}

\def\bi{\begin{itemize}} \def\ei{\end{itemize}}

\def\({\left(} \def\){\right)}
\def\[{\left[} \def\]{\right]}

\title{Classical 3-loop 2-body diagrams}

\author{Barak Kol and Ruth Shir \\
{\it Racah Institute of Physics, Hebrew University, Jerusalem 91904, Israel} \\
{\tt barak.kol,ruth.shir@mail.huji.ac.il}
}

\abstract{As part of the study of the two-body problem in Einstein's gravity, the fourth post-Newtonian order (4PN) of the two-body effective action is being computed presently by both effective field theory (EFT) methods and others. Diagrams with 3 (or 4) classical loops appear to be a significant obstacle. In this paper we develop a method to compute such 3-loop diagrams and demonstrate it through a specific diagram. We reduce the classical diagrams through shrinking the body worldlines to a form more familiar in Quantum Field Theory. A key ingredient in the evaluation is the Integration By Parts method for Feynman integrals.}

\begin{document}
\maketitle

\section{Introduction}

When considering the two-body problem in Einstein's gravity within the post-Newtonian limit (see the review \cite{BlanchetRev} and references therein) it is of central interest to compute the conservative two-body effective action $S_{2bd}[x_1,x_2]$ (we shall limit ourselves to non-spinning bodies) which in the effective field theory (EFT) approach \cite{GoldbergerRothstein1} is given by \be
S_{2bd}[x_1,x_2]:= \parbox{17mm}{\includegraphics[scale=0.4]{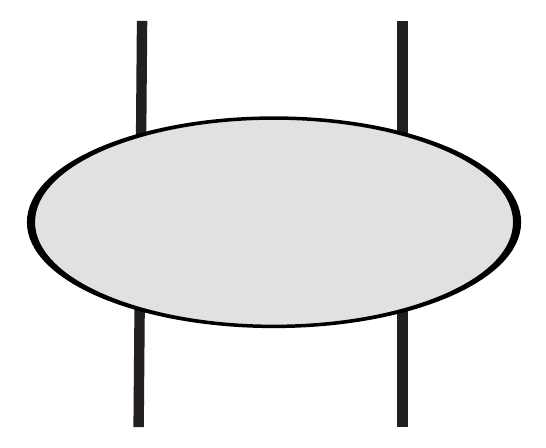}} \hspace{17mm}~, 
\label{def:S_2bd} \ee
where the heavy lines represent the two bodies and the oval represents all classical diagrams whose internal lines are non-relativistic gravitational fields \cite{CLEFT-caged,NRG}. 

Having reproduced $S_{2bd}$ within the EFT approach up to 3PN \cite{GoldbergerRothstein1,NRG,GilmoreRoss,FoffaSturani3PN} (assuming the harmonic gauge) an effort is presently underway to compute the yet unknown order 4PN. Certain sectors of 4PN were computed both in the EFT approach \cite{FoffaSturani4PNa,FoffaSturani4PNb} and in the Hamiltonian ADM formalism \cite{JaranowskiSchaefer4PNa}. Yet, computing diagrams with 3 classical loops appears to be a significant obstacle \cite{SturaniFoffa-private}. In this paper we shall develop a method to compute such diagrams, and quite generally classical 3-loop diagrams.

{\bf Motivation.} A good problem requires both experimental and theoretical motivation. The experimental motivation here is to contribute to the worldwide effort to detect gravitational waves, see for example \cite{IFOprojects}.  It is often said that for the purpose of constructing theoretical templates for a detector such as LIGO \cite{aLIGO} contributions up to order 3PN are necessary, see for example \cite{BlanchetRev}.%
\footnote{Paraphrasing from the top of p.8 of \cite{BlanchetRev} (rather than citing -- to avoid a possible automatic and erroneous ``text overlap" arXiv admin note): Measurement analyses has shown that 3PN precision ($1/c^6$ beyond the quadrupole moment approximation) is needed in order to find the best filtering technique in the LIGO and VIRGO detectors. This is because of the large number of orbital rotations that will be observed in the detectors frequency bandwidth which will allow to measure very accurately the orbital phase of the binary. The 3PN order is needed to compute the time evolution of the orbital phase.}
However, this is clearly an approximate statement depending on the parameters of both the binary system and the detector, and it cannot be excluded that 4PN effects would be measurable under some circumstances.

The theoretical motivation, which is more significant for us, will be revealed in two parts. First, when we shortly review the situation in lower orders and fewer loops we shall see that the first 2-loop diagram is also hard to compute yet its value is simple \cite{GilmoreRoss}.  That is often a clue for a hidden, simpler theory. Second, when we proceed to study the problem we will 
observe that the method known as Integration By Parts (IBP) \cite{ChetyrkinTkachov,SmirnovBook}, introduced to the EFT approach to GR in \cite{GilmoreRoss}, develops into a fascinating theory of classical diagrams.

\begin{figure}[t]
\centering
\includegraphics[width=10cm]{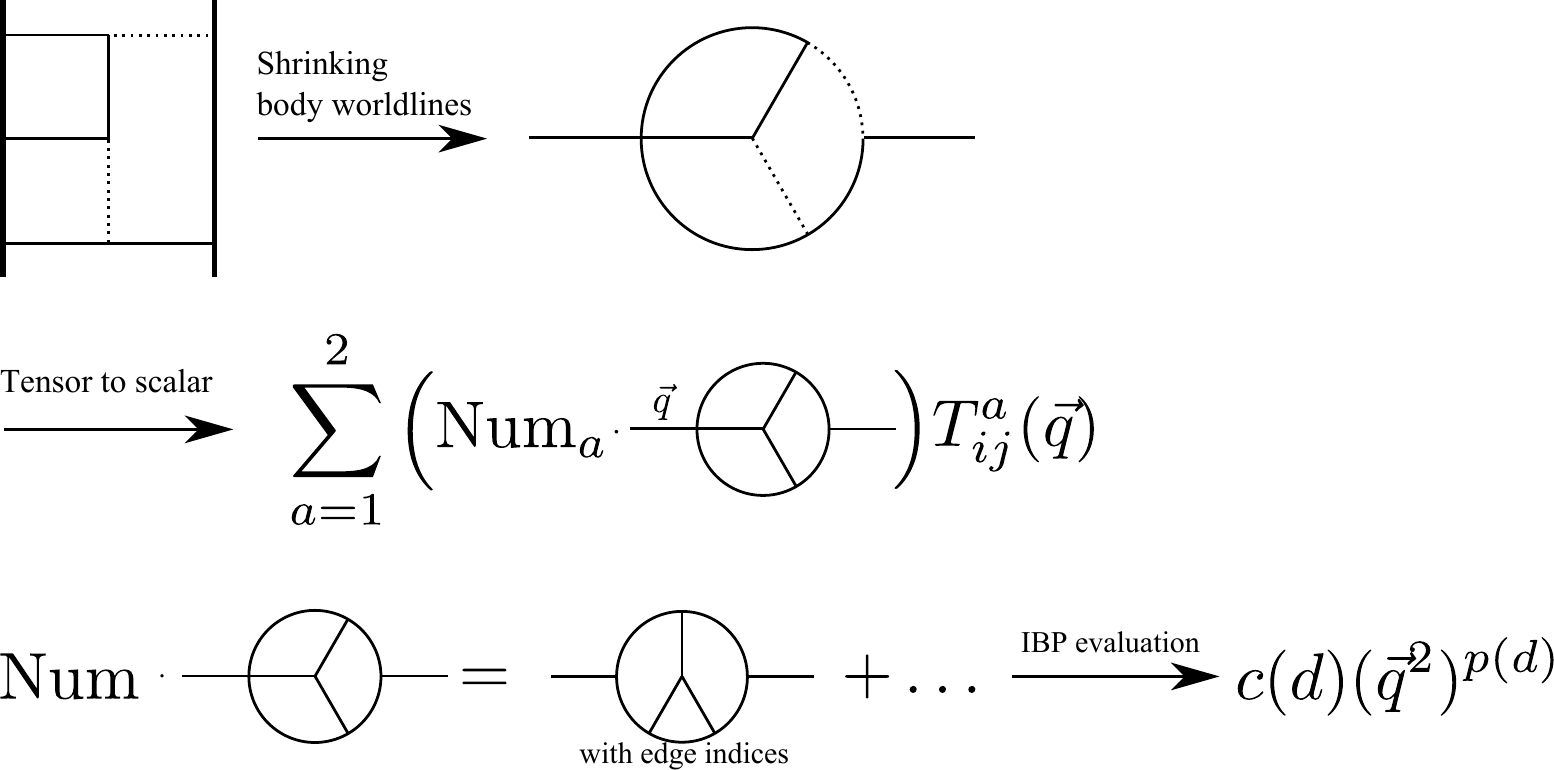}
\caption{An outline of the steps in the computation. $T_{ij}^a$ stands for the two possible tensor structures, and $\rm{Num}_a$ are two corresponding numerators. On general grounds, the final result contains only a dimension dependent constant, $c(d)$, and a known power $p(d)$.}
\label{fig:outline}
\end{figure}

This paper is organized as follows. In the remainder of the Introduction we discuss a 2-loop case as a warm up. Section \ref{sec:diag} introduces a specific diagram which we shall study, and we proceed to reduce it via a sequence of steps outlined in figure \ref{fig:outline}. In section \ref{sec:IBP} we describe the main computation within the IBP method. Finally in section \ref{sec:results} we summarize the results and discuss them.

\subsection*{Warm-up: Lower loops}

\begin{figure}[t]
\centering
\includegraphics{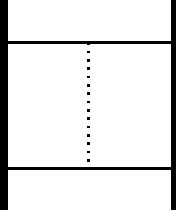}
\caption{The irreducible 2-loop diagram at order 2PN.}
\label{fig:2loop}
\end{figure}

The relevant action terms and Feynman rules are summarized in appendix \ref{app:FeynmanRules}. We recall that a classical (namely non-quantum) loop is a loop which some of its edges are non-propagating, that is they are the worldlines of the bodies. It is considered a loop because its computation involves standard loop integrals (as we shall also see explicitly in this paper).

No loops are required for order 1PN \cite{NRG} (as long as one uses NRG field variables). At 2PN there are both 1-loop diagrams and a 2-loop diagram, which were all computed in \cite{GilmoreRoss}. While the 1-loop diagrams can be computed in a straightforward manner, the 2-loop diagram, shown in figure \ref{fig:2loop} is irreducible in the sense of \cite{dressed}%
\footnote{Like all Feynman diagrams fig. \ref{fig:2loop} has associated stories. The leading Newtonian potential can be written as $\phi=\phi_1 + \phi_2$ where $\phi_a=-G\, m_a/|r-r_a|, ~ a=1,2$. The energy momentum tensor of $\phi$ is quadratic in $\phi$ and we shall consider the mixed term proportional to $\phi_1\, \phi_2$. This term sources the spatial metric $\sigma_{ij}$. Finally this diagram computes the energy stored in this mixed component of $\sigma_{ij}$.}
 and its computation requires an additional tool, which is the IBP method. Using it \cite{GilmoreRoss} found after evaluating and summing several terms \be
 \mbox{fig. } \ref{fig:2loop} = 2 \frac{G^3\, m_1^2\, m_2^2}{r^3}   ~.
 \ee
We note that all the dependence on parameters can be obtained from glancing at the diagram together with dimensional analysis, so the essential content of the computation is the dimensionless pre-factor $2$. Interestingly, in non-EFT methods all integrals up to 3.5PN (and hence at most 2-loops) are evaluated directly in configuration space.

In this case we were able to obtain an alternative derivation which does not use IBP, and instead takes place in the space of Schwinger parameters (we refer to its quotient by an overall scale as the Schwinger space). Schwinger space has the advantage of replacing several $d$ dimensional integrations over wave-numbers by several 1-dimensional Schwinger parameters. Once the overall scale of the Schwinger parameters is integrated we are left with an integral over the Schwinger space, a simplex $\Delta_n$ whose dimension $n$ is the number of propagators minus 1. Thus in the case at hand (figure \ref{fig:2loop}) the Schwinger space is $\Delta_4$. The identity (\ref{sigma-phi4}) is useful in working out the numerator of this diagram. In Schwinger space the integrand turns out to depend only on 2 linear combinations (out of the 4 coordinates on $\Delta_4$) allowing an immediate reduction to an integral over $\Delta_2$, which is simply a triangle. This last integral can be performed by integrating the variables ``one by one''. In this way we were able to replace the choices made in the IBP method and the various terms which appear there by a single straightforward method. In 3 loops we shall find that IBP enables to compute a typical diagram. However, so far we were not able to improve on the computation by working in Schwinger space.

\section{The diagram and its reduction}
\label{sec:diag}

\begin{figure}[t]
\centering
\includegraphics{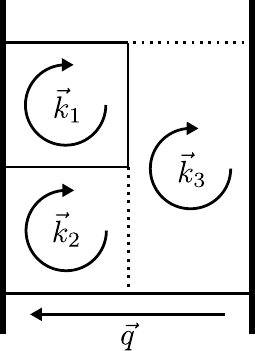}
\caption{The classical 3-loop diagram which we shall compute. The legend and Feynman rules are explained in appendix \ref{app:FeynmanRules}. Our choice of wavenumber variables is shown.}
\label{fig:pndiagram}
\end{figure}

For concreteness we chose to compute a specific and rather typical 3-loop diagram shown in figure \ref{fig:pndiagram}. Counting powers from worldline vertices confirms it to be 4PN. Later in subsection \ref{subsec:classif} we shall see that in some sense it has the most general topology for a classical 3-loop diagram. We proceed to describe a method to compute this diagram which we expect to apply to the whole class.

Assigning wavenumber loop variables  as shown in Figure \ref{fig:pndiagram} and using the Feynman rules from appendix \ref{app:FeynmanRules} the value of this diagram is given by \begin{eqnarray}
 \mbox{fig. } \ref{fig:pndiagram} = -C_0\,c_d^{-2}\, \mathbf{v}_{2i}\mathbf{v}_{2j}\int \frac{\mathrm{d}^d\mathbf{q}}{(2\pi)^d}e^{iqr}\int\frac{\mathrm{d}^d\mathbf{k}_1}{(2\pi)^d}\int\frac{\mathrm{d}^d\mathbf{k}_2}{(2\pi)^d}\int\frac{\mathrm{d}^d\mathbf{k}_3}{(2\pi)^d}\nonumber\\
\times \frac{F(\mathbf{k}_1,\mathbf{k}_2,\mathbf{k}_3,\mathbf{q})K_{ij}(\mathbf{k}_1,\mathbf{k}_3)}{D(\mathbf{k}_1,\mathbf{k}_2,\mathbf{k}_3,\mathbf{q})}
\end{eqnarray}
where \be 
 C_0=4096\pi^4 G^4 m_1^3 m_2^2 ~,
 \label{def:c0} \ee $c_d$ is defined in (\ref{def:cd}), the denominator is given by the product of propagators \be
D(\mathbf{k}_1,\mathbf{k}_2,\mathbf{k}_3,\mathbf{q}) := \mathbf{k}_1^2\mathbf{k}_3^2(\mathbf{k}_2-\mathbf{k}_1)^2(\mathbf{k}_3-\mathbf{k}_1)^2(\mathbf{k}_3-\mathbf{k}_2)^2(\mathbf{k}_2+\mathbf{q})^2(\mathbf{k}_3+\mathbf{q})^2 \ee
 and the terms in the numerator (where (\ref{sigma-phi4}) was useful) are given by \be
K_{ij}(\mathbf{k}_1,\mathbf{k}_2,\mathbf{k}_3,\mathbf{q}) := \mathbf{k}_{1i}\mathbf{k}_{3j}-\mathbf{k}_{1i}\mathbf{k}_{1j}
\end{equation}
 and  
\bea
 F(&\mathbf{k}_1&, \mathbf{k}_2,\mathbf{k}_3,\mathbf{q}) = \big((\mathbf{k}_2-\mathbf{k}_1)\cdot(\mathbf{k}_3+\mathbf{q})\big)\big((\mathbf{k}_3-\mathbf{k}_1)\cdot(\mathbf{k}_2+\mathbf{q})\big)  \non
 &+& \big((\mathbf{k}_2-\mathbf{k}_1)\cdot(\mathbf{k}_2+\mathbf{q})\big)\big((\mathbf{k}_3-\mathbf{k}_1)\cdot(\mathbf{k}_3+\mathbf{q})\big)  \non 
&-& 
\big((\mathbf{k}_2-\mathbf{k}_1)\cdot(\mathbf{k}_3-\mathbf{k}_1)\big)\big((\mathbf{k}_2+\mathbf{q})\cdot(\mathbf{k}_3+\mathbf{q})\big)  ~.
\eea
$d$ is the space dimension and it is kept general in anticipation of dimensional regularization.

\subsection{Shrinking worldlines and diagram topology classification}
\label{subsec:classif}

\begin{figure}[t]
\centering
\includegraphics[width=12cm]{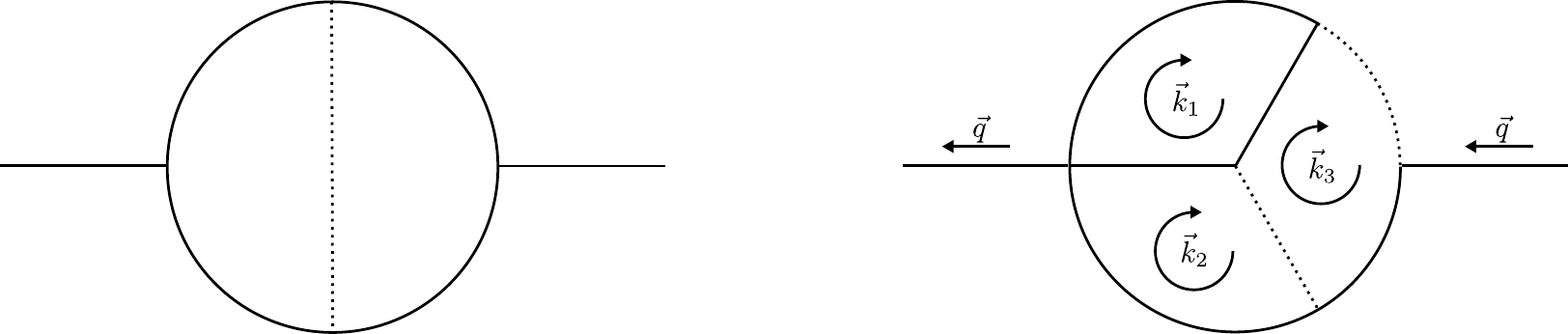}
\caption{The diagrams in fig. \ref{fig:2loop},\ref{fig:pndiagram} after shrinking the body worldlines.}
\label{fig:shrink}
\end{figure}

Since the body worldlines are non-propagating we can reduce the diagram by shrinking both worldlines to a point, and replacing the  worldline vertices by a single effective vertex for each body. As all diagrams in the 2-body problem are manifestly 2 point functions, we proceed to transform them to  wavenumber space, introducing $\mathbf{q}$ to be the wave transform of $\mathbf{r}_1-\mathbf{r}_2$. 
\footnote{These shrunk diagram were essentially pointed out by A. Ross to one of us (BK) at Jerusalem in December 2009.}
Fig. \ref{fig:shrink} shows the shrunk 2-loop and 3-loop diagrams. We note that in the new diagrams the classical loops appear quantum.

We can now classify the topology of irreducible classical 2-point 3-loop diagrams . By diagram topology we mean a diagram where lines are rendered indistinguishable (say solid) and hence the topology encodes the denominator, but not the numerator. By irreducible we mean that the diagram is dressing irreducible (it has no dressed sub-diagrams, see section 2.3 of \cite{dressed}) and it is bubble-free. A bubble (or parallel-loop) is a loop with 2 propagators, and can be evaluated immediately as discussed in subsection \ref{subsec:parallel}. 

\begin{figure}[t]
\centering
\includegraphics{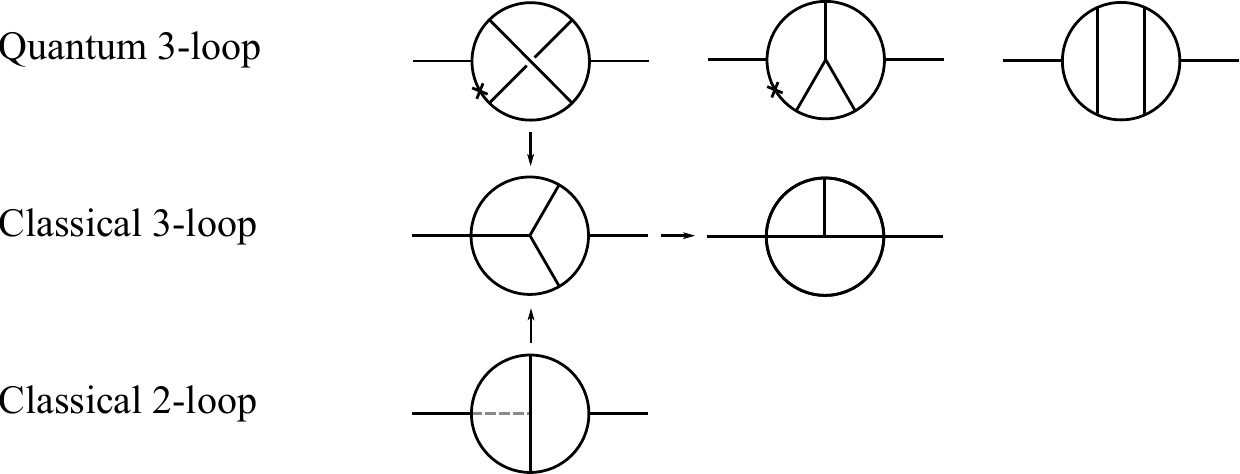} 
\caption{A topological classification of classical 2-body 3-loop diagrams. The dashed grey line transforms the 2-loop diagram to 3-loops.}
\label{fig:classif}
\end{figure}

The classification is shown in fig \ref{fig:classif}. The left side of the middle row shows the topology of our diagram (fig. \ref{fig:pndiagram}) with its 7 propagators and to its right is a 6-propagator degeneration. Altogether the middle row shows the two possible topologies. This classification can be obtained from two different directions. In the bottom line we show the (only) irreducible classical 2-loop diagram (all the diagrams are 2-point). It can be refined to a classical 3-loop diagram in a unique way, as shown by the dashed grey line. Alternatively, one can start with the three possible \emph{quantum} irreducible 3-loop diagram topologies (all having 8 propagators), shown on the top row, and consider their various classical degenerations -- a cross marks an edge to be shrunk. The top right diagram has no classical irreducible 3-loop degeneration.

\subsection{Simplifying the numerator}

{\bf Reducing tensor expression to scalars}. The tensor part in the numerator, $K_{ij}$, can be transformed into scalar components by realizing that the only tensor quantities on which the integral over the loop wavenumbers can depend are $\delta_{ij}$ and $\mathbf{q}_i\mathbf{q}_j$, and solving $\int_{\mathbf{k}_1}\int_{\mathbf{k}_2}\int_{\mathbf{k}_3}\frac{F(\mathbf{k},\mathbf{q})K_{ij}(\mathbf{k}_1,\mathbf{k}_3)}{D}=A(\mathbf{q})\delta_{ij}+B(\mathbf{q})\mathbf{q}_i\mathbf{q}_j$. In our case we separated the calculation into two parts, according to the terms appearing in the numerator:
\bea
\mbox{fig. } \ref{fig:pndiagram} &=& -C_0\,c_d^{-2} \frac{1}{d-1}\mathbf{v}_{2i}\mathbf{v}_{2j} 
\non &\times& \int_{\mathbf{q}}e^{i\mathbf{qr}}\Big[\Big(\delta_{ij}-\frac{\mathbf{q}_i\mathbf{q}_j}{\mathbf{q}^2}\Big)I_1(\mathbf{q})
+
\frac{1}{\mathbf{q}^2}\Big(-\delta_{ij}+d\frac{\mathbf{q}_i\mathbf{q}_j}{\mathbf{q}^2}\Big)I_2(\mathbf{q})\Big]
\eea
where
\bea
I_1(\mathbf{q}) &:=& \int_{\mathbf{k}_1}\int_{\mathbf{k}_2}\int_{\mathbf{k}_3}\frac{F(\mathbf{k},\mathbf{q})(\mathbf{k}_1\cdot\mathbf{k}_3-\mathbf{k}_1^2)}{D}
\non
I_2(\mathbf{q}) &:=& \int_{\mathbf{k}_1}\int_{\mathbf{k}_2}\int_{\mathbf{k}_3}\frac{F(\mathbf{k},\mathbf{q})\big((\mathbf{k}_1\cdot\mathbf{q})(\mathbf{k}_3\cdot\mathbf{q})-(\mathbf{k}_1\cdot\mathbf{q})^2\big)}{D}
\eea

\presub {\bf Expressing the numerator in terms of propagators as much as possible}. Following \cite{SmirnovBook} we proceed to express the numerator in terms of propagators as much as possible, by using $2\mathbf{k}_1\cdot\mathbf{k}_2=(\mathbf{k}_1+\mathbf{k}_2)^2-\mathbf{k}_1^2-\mathbf{k}_2^2$. In this way we simplify the numerator and either at the expense of introducing propagator indices $a_r$ defined such that the denominator is \be
 D := \Pi_r D_r^{a_r} ~, \ee 
where $D_r := \mathbf{E}_r^2$ are individual propagators. In some situations an index vanishes representing a simplification of the denominator as well. And even when it is non-vanishing it can be dealt with.  

In general not all terms in the numerator can be expressed in terms of propagators and one should use some additional invariants. In our case a single invariant sufficed (even though a count of invariants indicates  that two were possible {\it a priori}). Moreover, we found it to be possible to represent the additional invariant as a new propagator (with negative power) obtained through a resolution of the 4-vertex in the diagram into a pair of 3-vertices. However, it is not clear to us whether replacing the numerator invariant by the resolved diagram is essential to the method.

\begin{figure}[t]
\centering
\includegraphics[scale=0.5]{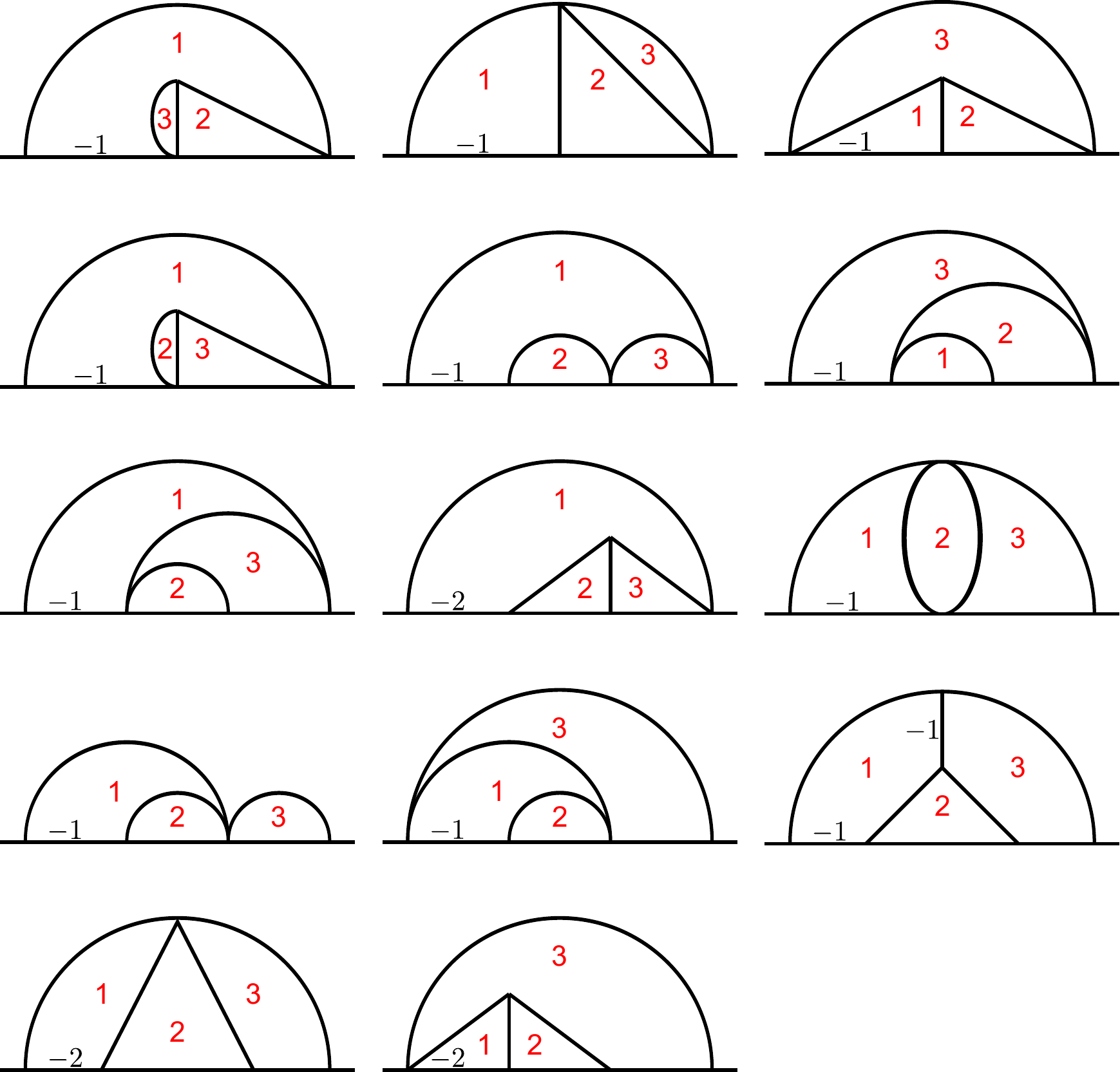}
\caption{The 14 auxiliary diagrams contributing to $I_1(\mathbf{q})$.}
\label{i1}
\end{figure}

We define an auxiliary diagram to be a representation of a Feynman integral such that the denominator is represented by indistinguishable lines, possibly with indices, and the numerator is represented either by negative index propagators or by its explicit expression. Loop integrations are read from the diagram as usual. 
We can now represent $I_1(\mathbf{q})$ and $I_2(\mathbf{q})$ as a sum of auxiliary diagrams. Figure \ref{i1} shows the auxiliary diagrams contributing to $I_1(\mathbf{q})$; the negative numbers are propagator indices and and the loops are numbered by their loop wavenumber variable. We found that 14 auxiliary diagrams contribute to $I_1$ and 43 auxiliary diagrams contribute to $I_2$.

\section{IBP evaluation}
\label{sec:IBP}

In this section we evaluate $I_1$ through the sum of auxiliary diagrams.

\subsection{Bubbles}
\label{subsec:parallel}

A bubble is replaced by a single line using the formula
\begin{equation}
\parbox{20mm}{\includegraphics[scale=0.6]{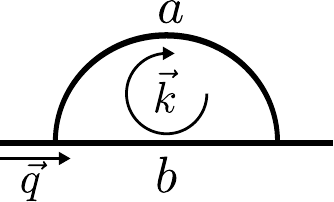}}
\equiv \int_{\mathbf{k}} \frac{1}{(\mathbf{k}^2)^a [(\mathbf{k}-\mathbf{q})^2]^b}=G(a,b,d)\, \frac{1}{(\mathbf{q}^2)^{a+b-d/2}}\label{pl} \equiv  
G(a,b,d)\, \parbox{20mm}{\includegraphics[scale=0.6]{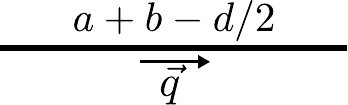}}
\end{equation}
where \be 
G(a,b,d) := \frac{1}{(4\pi)^{d/2}}\frac{\Gamma(a+b-d/2)\Gamma(d/2-a)\Gamma(d/2-b)}{\Gamma(a)\Gamma(b)\Gamma(d-a-b)} ~.
\label{def:G} \ee
A Generalization of this rule to the case with a numerator, see for example \cite{SmirnovBook}, is also useful 
\begin{equation}
\int_{\mathbf{k}}\frac{(2\mathbf{p}\cdot\mathbf{k})^n}{(\mathbf{k}^2)^a[(\mathbf{k}-\mathbf{q})^2]^b}=\frac{1}{(\mathbf{q}^2)^{a+b-d/2}} \sum_{r=0}^{[n/2]}F(a,b,d;r,n)\frac{n!}{r!(n-2r)!}(\mathbf{q}^2)^r(\mathbf{p}^2)^r(2\mathbf{q}\cdot\mathbf{p})^{n-2r}\label{pln}
\end{equation}
where \be
 F(a,b,d;r,n) := \frac{1}{(4\pi)^{d/2}} \frac{\Gamma(a+b-d/2-r)\, \Gamma(d/2-a+n-r)\, \Gamma(d/2-b+r)\,}{\Gamma(a)\, \Gamma(b)\, \Gamma(d-a-b+n)} ~.
 \label{def:F} \ee 
We note that sometimes one can evaluate the same integral more conveniently using the triangle rule (\ref{triangle-rule}) and viewing the numerator as a third loop edge with a negative index.
 
\subsection{IBP relations}
Some multi-loop Feynman integrals can be reduced to a sum of simpler integrals by applying the method of integration by parts (IBP). For a loop with momentum $\mathbf{k}$ and edges $\mathbf{E}_1^2,...,\mathbf{E}_n^2$ of the form $\mathbf{E}^2=(\mathbf{k}-\mathbf{p})^2$, the IBP relation \be
\int_{\mathbf{k}} \frac{\partial}{\partial\mathbf{k}} \mathbf{E} \frac{1}{(\mathbf{E}_1^2)^{a_1}(\mathbf{E}_2^2)^{a_2}...(\mathbf{E}_n^2)^{a_n}}=0
 \label{IBPrelation}
 \ee
 where $\mathbf{E}$ can be any wavenumber in the diagram, gives a sum of integrals in which one of the indices decreases while another increases. 
  When using IBP relations we specify both $\mathbf{k}$ and $\mathbf{E}$, and the latter is usually taken to be one of the edges of the loop $\mathbf{k}$.  
 
\subsubsection*{The Triangle Rule}
The triangle rule is a useful integration-by-parts relation which can be described diagrammatically \begin{eqnarray}
\raisebox{-1.5em}{\includegraphics[scale=0.6]{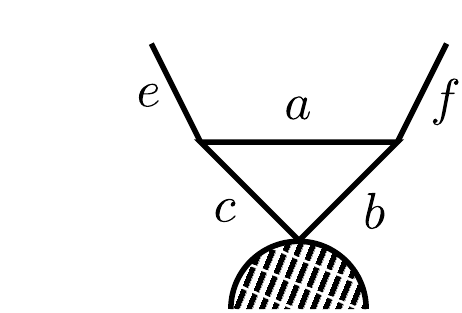}}=
\frac{1}{d-2a-b-c}\Big[b\Big(\raisebox{-1.5em}{ \includegraphics[scale=0.6]{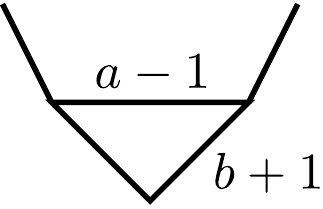}}-\raisebox{-1.5em}{\includegraphics[scale=0.6]{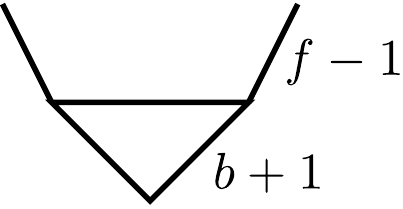}}\Big)\nonumber\\
+c\Big(\raisebox{-1.5em}{\includegraphics[scale=0.6]{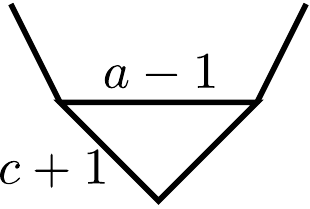}}-\raisebox{-1.5em}{\includegraphics[scale=0.6]{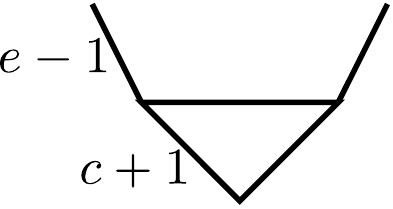}}\Big)\Big]
 \label{triangle-rule}
\end{eqnarray}

In our case it was also useful to apply a ``square rule'' for a loop with four edges 
\begin{eqnarray}
\raisebox{-1.5em}{\includegraphics[scale=0.6]{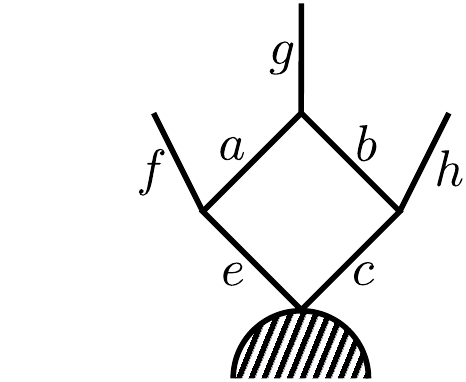}}=\frac{1}{d-2a-b-c-e}\Big[b\Big(\raisebox{-1.5em}{ \includegraphics[scale=0.6]{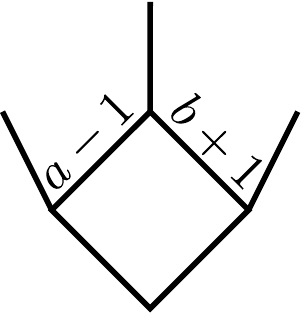}}-\raisebox{-1.5em}{\includegraphics[scale=0.6]{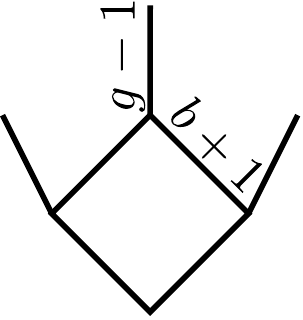}}\Big)\nonumber\\
+c\Big(\raisebox{-1.5em}{\includegraphics[scale=0.6]{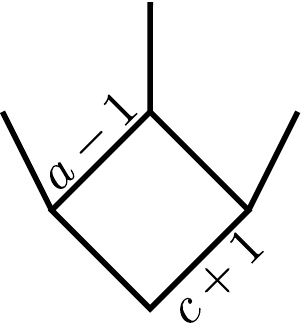}}-\raisebox{-1.5em}{\includegraphics[scale=0.6]{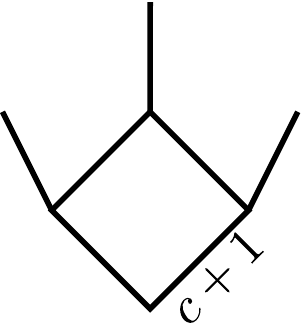}}(\mathbf{E}_a-\mathbf{E}_c)^2\Big)\nonumber\\
+e\Big(\raisebox{-1.5em}{ \includegraphics[scale=0.6]{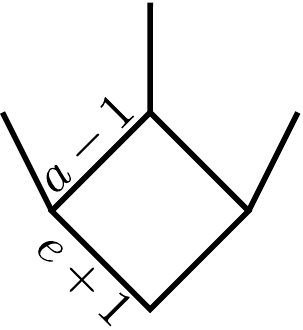}}-\raisebox{-1.5em}{\includegraphics[scale=0.6]{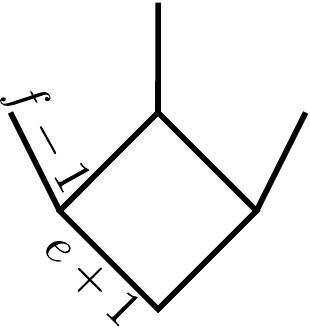}}\Big)\Big]
\end{eqnarray}
where $\mathbf{E}_a$ and $\mathbf{E}_c$ are the edges corresponding to $a$ and $c$.

\subsection{Calculating auxiliary Diagrams} 
In some of the auxiliary diagrams it is sufficient to use the bubble formula repeatedly. In others we had to use IBP relations such as the triangle rule (\ref{triangle-rule}). The IBP relations may be iterated a few times until we reach diagrams which contain bubbles only. 
One of the hardest diagrams in $I_2(\mathbf{q})$ is the following
\begin{equation}
\raisebox{-1.5em}{\includegraphics[scale=0.5]{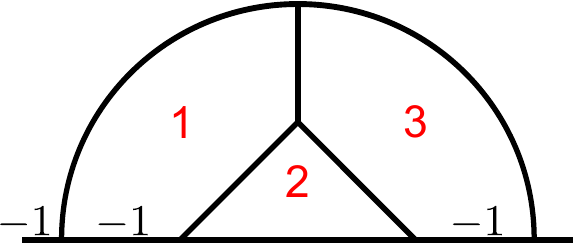}}=\int_{\mathbf{k}_1}\int_{\mathbf{k}_2}\int_{\mathbf{k}_3}\frac{\mathbf{q}^2(\mathbf{k}_1+\mathbf{q})^2(\mathbf{k}_3+\mathbf{q})^2}{\mathbf{k}_1^2\mathbf{k}_3^2(\mathbf{k}_2-\mathbf{k}_1)^2(\mathbf{k}_3-\mathbf{k}_1)^2(\mathbf{k}_3-\mathbf{k}_2)^2(\mathbf{k}_2+\mathbf{q})^2}
\end{equation}

By applying the square rule to loop 3 and edge $\mathbf{k}_3-\mathbf{k}_1$ we reduce the diagram to the following sum of simpler diagrams:
 
\begin{eqnarray}
\frac{1}{d-3}\Big(\raisebox{-1.5em}{\includegraphics[scale=0.5]{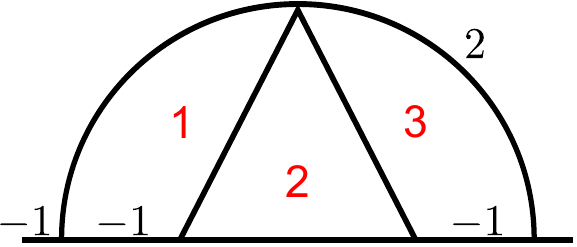}}-\raisebox{-1.5em}{\includegraphics[scale=0.5]{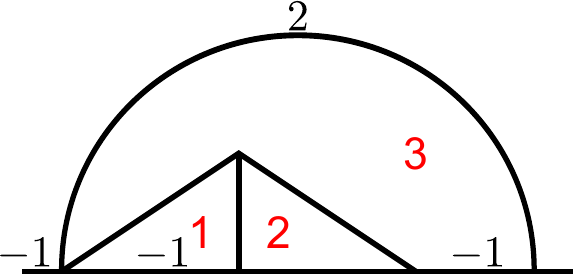}}\nonumber\\
-\raisebox{-1.5em}{\includegraphics[scale=0.5]{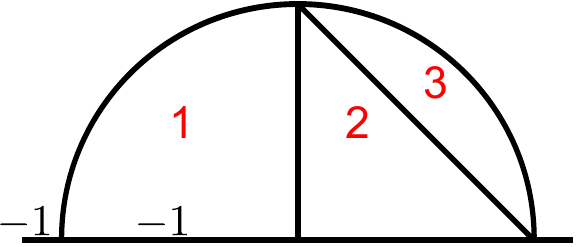}}+\raisebox{-1.5em}{\includegraphics[scale=0.5]{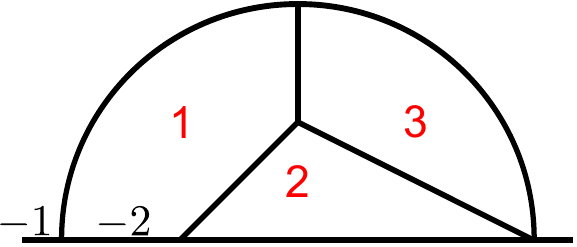}}\nonumber\\
+\raisebox{-1.5em}{\includegraphics[scale=0.5]{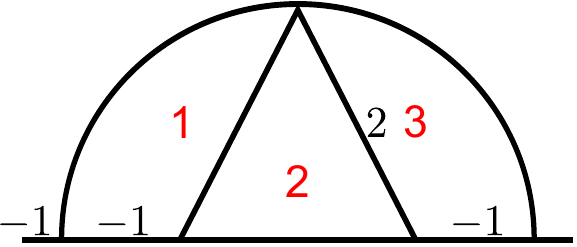}}-\raisebox{-1.5em}{\includegraphics[scale=0.5]{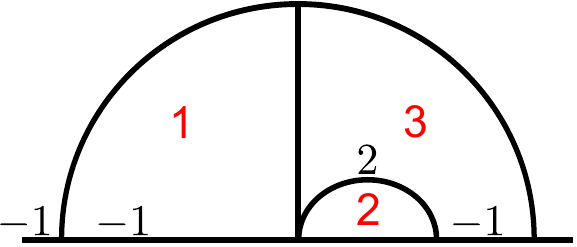}}\Big)
\end{eqnarray}

Each of these diagrams can be calculated by applying further IBP relations and/or equations (\ref{pl}, \ref{pln}).  
The value of this auxiliary diagram turns out to be \be
-\frac{1}{1024}+\frac{1}{128\pi^2} ~. 
\ee

Evaluating in this manner $I_1$ and $I_2$ through all of their auxiliary diagrams we found \be 
12288\,\pi^2\, I_1(\mathbf{q}) = \big(-\frac{4}{\epsilon}+3\pi^2-12-6\gamma+16\log 2+6\log \pi+O(\epsilon)\big)(\mathbf{q}^2)^{\frac{1+3\epsilon}{2}} ~.
\label{I1a}
\ee
\be 
2457600\,\pi^2\, I_2(\mathbf{q}) = \big(-\frac{320}{\epsilon}+225\pi^2-648-480\gamma+1280\log 2+480\log \pi+O(\epsilon)\big)(\mathbf{q}^2)^{\frac{3+3\epsilon}{2}} ~.
\label{I2a}
\ee
where $\epsilon=d-3$ is the parameter of dimensional regularization which was required, and $\gamma$ is Euler's constant.
 The next step is to Fourier transform over $\mathbf{q}$ (see appendix \ref{app:Fourier}). We find that \bea
 \mbox{fig. } \ref{fig:pndiagram} &=& -C_0\,c_d^{-2}\,\frac{1}{\epsilon+2}\, \mathbf{v}_{2i}\mathbf{v}_{2j}\int_{\mathbf{q}}e^{i\mathbf{qr}}\Big[\big(\delta_{ij}-\frac{\mathbf{q}_i\mathbf{q}_j}{\mathbf{q}^2}\big)I_1(\mathbf{q})
+\frac{1}{\mathbf{q}^2}\Big(-\delta_{ij}+(\epsilon+3)\frac{\mathbf{q}_i\mathbf{q}_j}{\mathbf{q}^2}\Big)I_2(\mathbf{q})\Big] \non 
 &=& 
\frac{4}{15}\big(-\frac{1}{\epsilon}+2\gamma+2\log 4\pi \mathbf{r}^2\big)\frac{\mathbf{v}^2\mathbf{r}^2+2(\mathbf{v}\cdot\mathbf{r})^2}{\mathbf{r}^6}
\non &+&
\big(-\frac{106}{75}+\frac{\pi^2}{4}\big)\frac{(\mathbf{v}\cdot\mathbf{r})^2}{\mathbf{r}^6}+\big(-\frac{86}{25}+\frac{\pi^2}{4}\big)\frac{\mathbf{v}^2}{\mathbf{r}^4}
~.
 \label{I1b}
 \eea

\section{Summary of results and discussion}
\label{sec:results}

In this paper we developed a method to compute classical 3-loop 2-point diagrams such as those appearing at order 4PN of the conservative two-body effective action. The method consists of the following steps \begin{enumerate}
\item Shrink body worldlines
\item Decompose a tensor Feynman integral into a sum of scalar integrals
\item Express the numerator in terms of propagators as much as possible
\item Evaluate the resulting sum of ``auxiliary diagrams'' with the help of the IBP method.
\end{enumerate}
Apart from the first step this is essentially the method of Integration By Parts (IBP) \cite{ChetyrkinTkachov,SmirnovBook}  applied to a classical diagram.

We demonstrated the method by applying it to both $I_1$ and $I_2$, the 2 scalar components of the diagram in fig. \ref{fig:pndiagram}, performing all the steps explicitly and reaching the result in (\ref{I1a}, \ref{I2a}, \ref{I1b}). While we strove to calculate carefully we note that we have no independent check for this expression. Actually we view the method developed here to be the main result of the paper and the evaluation serves mainly to demonstrate that the method is complete.

The 2PN case led us to expect a simple result for a complicated  computation. The existence of a dimensional regularization pole makes the answer a bit more complicated. Still, the coefficient of the pole is indeed very simple. In addition we find that {\it a posteriori} the evaluation involved fascinating theoretical ideas.

{\bf The physical interpretation of the pole}. The pole must be cancelled by a counter-term of the form \be
\parbox{20mm}{\includegraphics[scale=0.7]{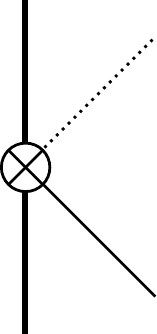}}
 ~. \ee
More generally 3PN counter-terms are known to arise as an artifact of the harmonic gauge. Indeed the form of the Schwarzschild metric in harmonic coordinates contains a $\log(r)$ term at order $(m/r)^3$ which is related to the 3PN counter-term \be  
 \parbox{20mm}{\includegraphics[scale=0.7]{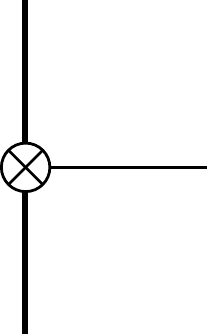}}
 ~. \ee

\presub {\bf Discussion}. 

{\bf Automatization and application to 4PN}. Application to 4PN would benefit from automatization of the procedure. This is presumably possible since IBP is commonly used in computerized computations (see for example \cite{HennSmirnovs}). Still, there are specific points that required our attention: step 3 depends on the choice of additional invariants (beyond the propagators); and in step 4 we applied human judgement in the course of choosing the loop and the edge to appear in the IBP relation (\ref{IBPrelation}). Finally, in addition to automatization order 4PN would require to evaluate also classical 4-loop diagrams, either by generalization of the current method or otherwise.

{\bf The IBP method}. Proceeding to discuss the theory of multi-loop computation, the IBP method presents a clear challenge, namely to characterize which diagrams are computable by IBP (see \cite{Baikov}  for results in this direction). Moreover, its very name is sub-optimal. Indeed the main IBP identity (\ref{IBPrelation}) is derived through integration by parts which is an elementary property of integrals. However, in order for the reduction to be useful the loop and edge variables must be chosen judiciously  in an attempt that the numerator will not get supplemented by new invariants. For this reason the method is strongly dependent on the diagram's topology, and hence its final formulation is expected to contain more ingredients beyond a mere integration by parts. 
 
{\bf Another method?}  In the 2-loop example we noted that we were able to avoid the arbitrariness and long computation associated with the IBP method by working in Schwinger space. Still, so far we were not able to generalize this idea to 3-loops.

\subsection*{Acknowledgments}

BK thanks the organizers of the conference ``Effective Field Theory and Gravitational Physics'', November 28 - 30, 2011 at the Perimeter Institute which contributed to this work.

This research was supported by the Israel Science Foundation grant no. 812/11.

\appendix

\section{Action and Feynman rules}
\label{app:FeynmanRules}

In this appendix we summarize the action terms and Feynman rules which we shall need. We work with non-relativistic gravitational fields (NRG fields), which are a redefinition of the Einstein field $g_{\mu\nu}$ in terms of the fields $\phi,\, \mathbf{A}_i,\, \sigma_{ij}$ given by \cite{CLEFT-caged,NRG} \be
 ds^2 = e^{2\phi} \( dt -2\, \mathbf{A}_i\, d\mathbf{x}^i \)^2 - e^{-2 \phi/\hd} \gamma_{ij} d\mathbf{x}^i\, d\mathbf{x}^j ~,  \label{def:NRG}
 \ee
where $\phi$ is the Newtonian potential field, $\mathbf{A}_i$ is the gravito-magnetic vector potential (using the normalization conventions of \cite{PNRR}), $\gamma_{ij} \equiv \delta_{ij}+ \sigma_{ij}$ is the spatial metric and $\hd:=d-2$ where $d$ is the space dimension, and it is kept general in anticipation of dimensional regularization.

The total action is given by \be
 S = S_{EH} + S_{GF} + \sum_{A=1}^{2} S_A\[ x_A,\dots ;g_{\mu\nu}\] \label{def:S} ~. \ee
 For the purpose of the present paper we shall only use the stationary sector of $S_{EH}$, the Einstein-Hilbert action, re-expressed in terms of NRG fields \be
S_{EH} \(\phi,\, \mathbf{A}_i,\, \sigma_{ij}\)= \frac{1}{8 \pi G} \int \sqrt{\gamma}\, d^3\mathbf{x}\, dt \[ -c_d\, \left| \del_i \phi \right|^2 + e^{4 c_d\, \phi}  \mathbf{B}^2 + \frac{1}{2} R[\gamma] + {\cal O} \( \del_t \) \]  ~,
\label{def:S-EH} \ee
 where $\left| \del_i \phi \right|^2 := \gamma^{ij}\, \del_i \phi\, \del_j \phi$,  $\mathbf{B}^i:= \sqrt{\gamma}^{-1} \eps^{ijk}\, \del_j\, \mathbf{A}_k$ is the gravito-magnetic field strength (in arbitrary $d$ replace $\mathbf{B}^2 \to -\half F^2$) and the dimensional dependence is contained in%
\footnote{This definition is the same as that of \cite{FoffaSturani3PN} apart for a normalization $c_d^{\mbox{here}} = c_d^{[6]}/4$.} 
 \be
 c_d :=  \frac{\hd+1}{2 \hd} 
 \label{def:cd} \ee
 and hence $c_{d=3}=1$.
 The gauge fixing action is \be
 S_{GF}  = \frac{1}{8 \pi G} \int \sqrt{\gamma}\, d^3\mathbf{x}\, dt  \[ e^{4 c_d\, \phi} \( D^i \mathbf{A}_i \)^2  - \frac{1}{4} \left| \Gamma^i[\gamma] \right|^2 + {\cal O} \( \del_t \) \] ~.
 \label{def:S-GF} \ee
 where $\Gamma^i = \Gamma^i_{jk}[\gamma] \gamma^{jk}$ and $\Gamma^i_{jk}$ are the Christoffel symbols.
The full gauge-fixed action including time dependent terms and the dependence on the space-time dimension were obtained in \cite{NRGaction}. For the body action it will suffice to consider the point particle approximation \be
 S\[x;g_{\mu\nu}\] = -m \int d\tau = -m \int dt \sqrt{e^{2\phi} \(1-2\, \mathbf{A} \cdot \mathbf{v}\)^2 - e^{-2\phi/ \hd}\, \gamma_{ij}\, \mathbf{v}^i \mathbf{v}^j} ~,
 \label{def:Spp} \ee
 where $\mathbf{v}^i := d\mathbf{x}^i/dt$ is the 3-velocity.

We shall require only the following Feynman rules, all arising from the action above (see also \cite{dressed} for example) \bea
 \parbox{30mm}{\includegraphics[scale=0.4]{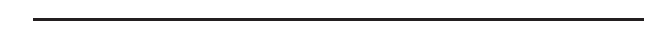}}
 &=& c_d^{-1}\,  \frac{4 \pi\, G} {\mathbf{k}^2}\, \delta(t_1-t_2) \non
\parbox{30mm}{\includegraphics[scale=0.4]{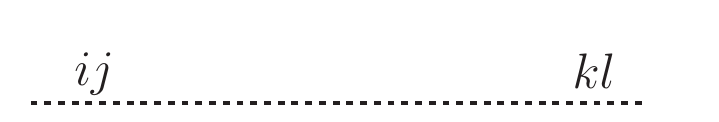}} 
&=& \frac{32 \pi\, G\, } {\mathbf{k}^2}\, \delta(t_1-t_2)\, P_{ij,kl} \qquad P_{ij,kl} := \half\( \delta_{ik}\, \delta_{jl} + \delta_{il}\, \delta_{jk} - \frac{2}{\hd}\, \delta_{ij} \delta_{kl} \) \non
\parbox{20mm}{\includegraphics[scale=0.7]{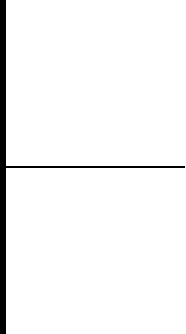}} 
&=& -m \int dt \non
\parbox{20mm}{\includegraphics[scale=0.7]{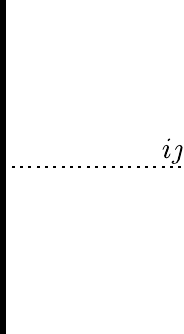}} 
 &=& \frac{m}{2} \int dt\, \mathbf{v}^i \mathbf{v}^j \non
\parbox{20mm}{\includegraphics[scale=0.7]{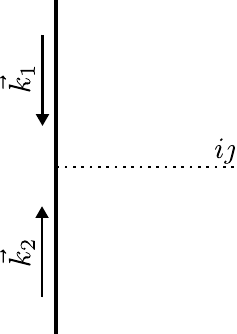}} 
  &=& -c_d\, \frac{1}{8 \pi G} \int dt \( \mathbf{k}_i\, \mathbf{q}_j + \mathbf{k}_j\, \mathbf{q}_i - ({\bf k} \cdot {\bf q})\, \delta_{ij} \) 
 \eea
 where we used real and $\hbar$-free Feynman rules conventions \cite{CLEFT-caged}.
 
 A useful composite expression is given by \be
\parbox{15mm}{\includegraphics[scale=0.8]{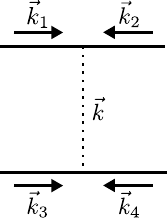}}
 = \int dt \int_\mathbf{k} \frac{c_d^2}{\pi G\, \mathbf{k}^2} \( (\mathbf{k}_1 \cdot \mathbf{k}_3) (\mathbf{k}_2 \cdot \mathbf{k}_4)  + (\mathbf{k}_1 \cdot \mathbf{k}_4) (\mathbf{k}_2 \cdot \mathbf{k}_3) - (\mathbf{k}_1 \cdot \mathbf{k}_2) (\mathbf{k}_3 \cdot \mathbf{k}_4) \)\, 
\label{sigma-phi4}
\ee
where $\mathbf{k}^2 = (\mathbf{k}_1+\mathbf{k}_2)^2 = (\mathbf{k}_3+\mathbf{k}_4)^2$.\\

\section{Fourier transform integrals}
\label{app:Fourier}

The $d$ dimensional Fourier transform integrals we have used to compute the final integral over $\mathbf{q}$ are:
\begin{equation}
\int_{\mathbf{q}}(\mathbf{q}^2)^{-a}e^{i\mathbf{qr}}=\frac{1}{(4\pi)^{d/2}}\frac{\Gamma(d/2-a)}{\Gamma(a)}\Big(\frac{4}{\mathbf{r}^2}\Big)^{d/2-a}
\end{equation}
\begin{equation}
\int_{\mathbf{q}}\mathbf{q}_i\mathbf{q}_j(\mathbf{q}^2)^{-a}e^{i\mathbf{qr}}=\frac{1}{(4\pi)^{d/2}}\frac{\Gamma(d/2-a+1)}{\Gamma(a)}\Big(\frac{\delta_{ij}}{2}+(a-d/2-1)\frac{\mathbf{r}_i\mathbf{r}_j}{\mathbf{r}^2}\Big)\Big(\frac{4}{\mathbf{r}^2}\Big)^{d/2-a+1}
\end{equation}



\begin{thebibliography}{99}

\bibitem{BlanchetRev}
  L.~Blanchet,
  ``Gravitational radiation from post-Newtonian sources and inspiralling compact binaries,''
  Living Rev.\ Rel.\ {\bf 5}, 3 (2002), update: Living Rev.\ Rel.\ {\bf 9}, 4 (2006)  [arXiv:gr-qc/0202016].
L.~Blanchet,
  ``Post-Newtonian theory and the two-body problem,''
  Fundam.\ Theor.\ Phys.\  {\bf 162}, 125 (2011)
  [arXiv:0907.3596 [gr-qc]].

\bibitem{GoldbergerRothstein1}
  W.~D.~Goldberger and I.~Z.~Rothstein,
 ``An effective field theory of gravity for extended objects,''
  Phys.\ Rev.\  D {\bf 73}, 104029 (2006). \\
  W.~D.~Goldberger,
  ``Les Houches lectures on effective field theories and gravitational radiation,''
  arXiv:hep-ph/0701129.
  
\bibitem{CLEFT-caged}
  B.~Kol and M.~Smolkin,
  ``Classical Effective Field Theory and Caged Black Holes,''
  Phys.\ Rev.\  D {\bf 77}, 064033 (2008)
  [arXiv:0712.2822 [hep-th]].

\bibitem{NRG}
  B.~Kol and M.~Smolkin,
  ``Non-Relativistic Gravitation: From Newton to Einstein and Back,''
  Class.\ Quant.\ Grav.\  {\bf 25}, 145011 (2008)
  [arXiv:0712.4116 [hep-th]].

\bibitem{GilmoreRoss}
  J.~B.~Gilmore and A.~Ross,
  ``Effective field theory calculation of second post-Newtonian binary dynamics,''
  Phys.\ Rev.\ D {\bf 78}, 124021 (2008)
  [arXiv:0810.1328 [gr-qc]].
    
\bibitem{FoffaSturani3PN} 
  S.~Foffa and R.~Sturani,
  ``Effective field theory calculation of conservative binary dynamics at third post-Newtonian order,''
  Phys.\ Rev.\ D {\bf 84}, 044031 (2011)
  [arXiv:1104.1122 [gr-qc]].
    
\bibitem{FoffaSturani4PNa}
  S.~Foffa and R.~Sturani,
  ``Tail terms in gravitational radiation reaction via effective field theory,''
  Phys.\ Rev.\ D {\bf 87}, 044056 (2013)
  [arXiv:1111.5488 [gr-qc]].
  
 \bibitem{FoffaSturani4PNb}
  S.~Foffa and R.~Sturani,
  ``The dynamics of the gravitational two-body problem in the post-Newtonian approximation at quadratic order in the Newton's constant,''
  Phys.\ Rev.\ D {\bf 87}, 064011 (2013)
  [arXiv:1206.7087 [gr-qc]].

\bibitem{JaranowskiSchaefer4PNa}
  P.~Jaranowski and G.~Schafer,
  ``Towards the 4th post-Newtonian Hamiltonian for two-point-mass systems,''
  Phys.\ Rev.\ D {\bf 86} (2012) 061503
  [arXiv:1207.5448 [gr-qc]].\\
  ``Dimensional regularization of local singularities in the 4th post-Newtonian two-point-mass Hamiltonian,''
  Phys.\ Rev.\ D {\bf 87} (2013) 081503
  [arXiv:1303.3225 [gr-qc]].
    
 \bibitem{SturaniFoffa-private}
S.~Foffa and R.~Sturani, private communication and a talk ``Conservative binary dynamics at 3PN order and beyond via effective field theory methods'',
 given at ``Effective Field Theory and Gravitational Physics'', November 28 - 30, 2011 Perimeter Institute. 
    
\bibitem{IFOprojects}
http://www.ligo.org/partners.php and  http://www.geo600.org/links/GWlinks/interferometric-detectors

\bibitem{aLIGO}
  S.~J.~Waldman [LIGO Scientific Collaboration],
  ``The Advanced LIGO Gravitational Wave Detector,''
  arXiv:1103.2728 [gr-qc].
  G.~M.~Harry [LIGO Scientific Collaboration],
  ``Advanced LIGO: The next generation of gravitational wave detectors,''
  Class.\ Quant.\ Grav.\  {\bf 27}, 084006 (2010).
      
\bibitem{ChetyrkinTkachov} 
  K.~G.~Chetyrkin and F.~V.~Tkachov,
  ``Integration by Parts: The Algorithm to Calculate beta Functions in 4 Loops,''
  Nucl.\ Phys.\ B {\bf 192}, 159 (1981).

\bibitem{SmirnovBook} 
  V.~A.~Smirnov,
  ``Feynman integral calculus,''
  Berlin, Germany: Springer (2006).

\bibitem{dressed} 
  B.~Kol and M.~Smolkin,
  ``Dressing the Post-Newtonian two-body problem and Classical Effective Field Theory,''
  Phys.\ Rev.\ D {\bf 80}, 124044 (2009)
  [arXiv:0910.5222 [hep-th]].

\bibitem{HennSmirnovs} 
  J.~M.~Henn, A.~V.~Smirnov and V.~A.~Smirnov,
  ``Analytic results for planar three-loop four-point integrals from a Knizhnik-Zamolodchikov equation,''
  arXiv:1306.2799 [hep-th]. See references [9-14].

\bibitem{Baikov} 
  P.~A.~Baikov,
  ``Explicit solutions of the three loop vacuum integral recurrence relations,''
  Phys.\ Lett.\ B {\bf 385}, 404 (1996)
  [hep-ph/9603267]. \\
  ``Explicit solutions of the multiloop integral recurrence relations and its application,''
  Nucl.\ Instrum.\ Meth.\ A {\bf 389}, 347 (1997)
  [hep-ph/9611449].

\bibitem{PNRR} 
  O.~Birnholtz, S.~Hadar and B.~Kol,
  ``A theory of post-Newtonian radiation and reaction,''
  arXiv:1305.6930 [hep-th].
  
\bibitem{NRGaction} 
  B.~Kol and M.~Smolkin,
  ``Einstein's action and the harmonic gauge in terms of Newtonian fields,''
  Phys.\ Rev.\ D {\bf 85}, 044029 (2012)
  [arXiv:1009.1876 [hep-th]].

\end{thebibliography}
\end{document}